\documentclass[a4paper,11pt]{article}

\pdfoutput=1 

\usepackage{jinstpub} 

\title{\boldmath EIC Detector Overview}

\author{Douglas W. Higinbotham}
\affiliation{Thomas Jefferson National Accelerator Facility, Newport News, VA, 23601, USA}

\emailAdd{doug@jlab.org}

\abstract{The Electron Ion Collider will have two interaction regions that can be instrumented with detectors.   The first region will be instrumented as part of the project and needs to be capable of delivering the physics that has been outlined by the National Academy of Sciences and ready at the start of beam commissioning near the end of this decade.   Plans for a second complementary detector to be located at a second interaction region are already in progress and will hopefully come to fruition just a few years after the first detector comes online.   While the basic parameters of these detectors are being selected using conventional approaches, the optimization of the detectors is already being enhanced by making use of advanced optimization techniques.}


\proceeding{Artificial Intelligence for the Electron Ion Collider\\
  7-11 September 2021}

\begin{document}
\maketitle
\flushbottom

\section{Introduction}
\label{introduction}

The science case has been made for the future Electron Ion Collider as detailed in the National Academy of Sciences report on the future EIC~\cite{NAP25171} as well as detailed in the EIC White Paper~\cite{Accardi:2012qut} and the
EIC yellow report~\cite{AbdulKhalek:2021gbh}.
The physics case includes, but is not limited to, determining how the nucleon's properties such as mass and spin emerge from partons, how partons are distributed in the nucleon, and how does the nuclear environment affect the dynamics of the guarks and gluons.
In order to experimentally be able to address these topics, the EIC needs to be built with an extremely versatile hermetic detector able to operate over a very wide region of particle momenta.   Design of this detector has mostly followed a conventional path with teams of scientists working on determining the general detector configurations, though already AI and machine learning techniques are being employed to refine those locations and optimize the size of the systems to get the best possible performance while also keeping costs under control.

\section{First Interaction Region}
\label{detectors}

The Electron Ion Colider project includes funding for one interaction region and one detector.  At a collider, the interaction region and the detector are tightly coupled and need to be designed with a close collaboration between the accelerator physicists and the nuclear physicists.    This ensures both that detector will preform as intended and that the beam will have the desired characteristics.    As an example, as the detector becomes longer, the focusing quadrupole magnets of the accelerator must be located further apart causing a trade-off between detector performance and maximum luminosity.

\begin{figure}[tp]
\centering 
\includegraphics[width=\linewidth]{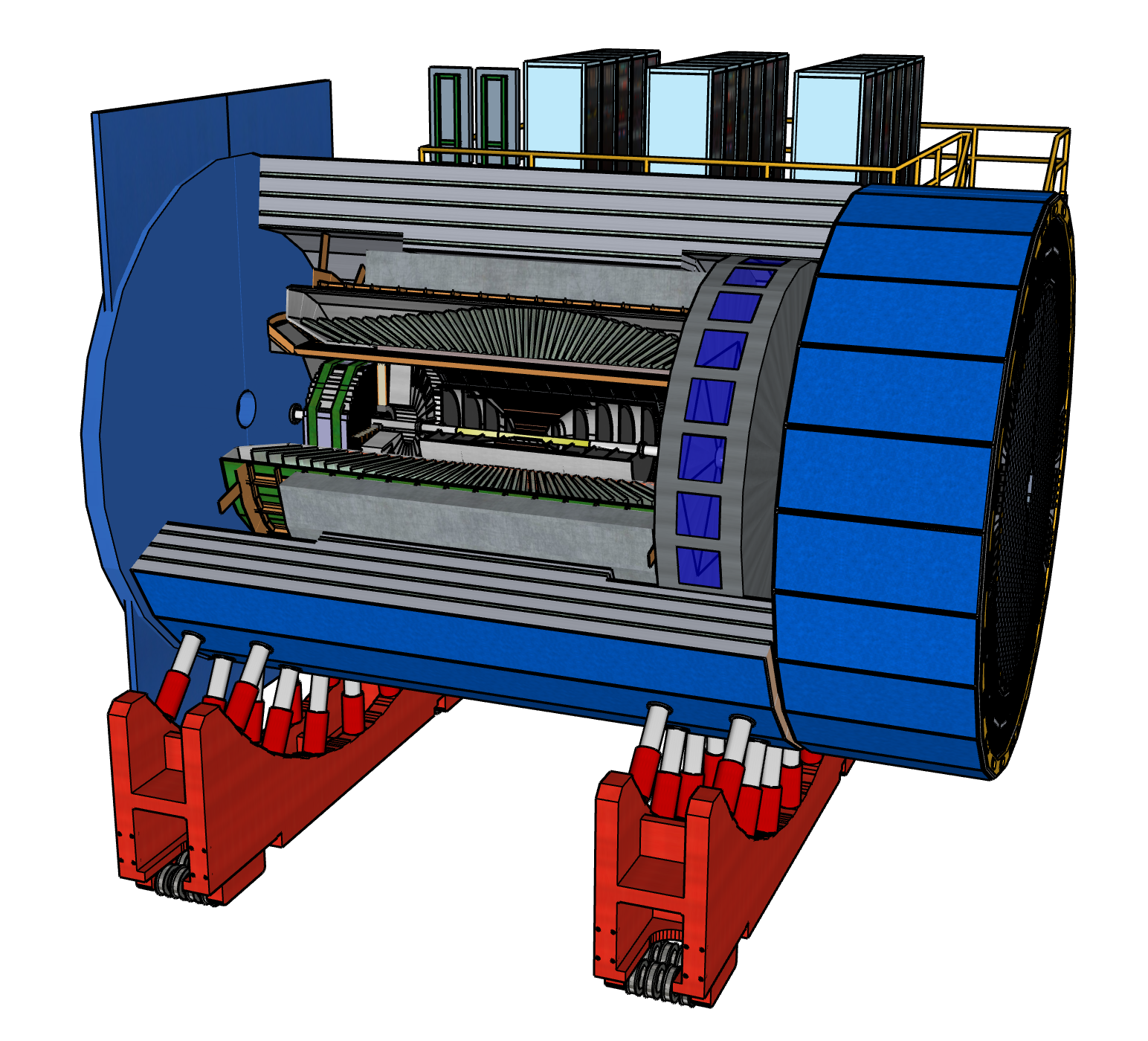}
\caption{\label{fig:ecce} An image of one of the Electron Ion Collider detector concepts as detailed in SketchUp Pro software.   In this image, the ECCE detector concept can be seen on its cradle and with the three levels of associated electronics located behind it.   A complete database of the various detector concepts for ATHENA, CORE and ECCE detectors can be found online~\cite{Akers:2020}.}
\end{figure}

There are currently three competing ideas for the first detector by the ATHENA, CORE and ECCE proto-collaborations.
All three designs make use of a solenoid magnet with either a new 3~Tesla magnet or reuse of the BaBar 1.5~Tesla magnet~\cite{Fabbricatore:1996mc}.
All three designs are working to be able to deliver the science of the National Academy report.   
In general, the basics characteristics of these detectors are the same, though there are a great deal of differences in the details.    All the detector systems will be comprised of  silicon tracking detectors~\cite{Arrington:2021yeb}. Particle identification detectors
such as 
a dual ring imaging Cherenkov detector (dRICH)~\cite{Cisbani:2019xta}, modular RICH~\cite{Agarwala:2019wcy}, and
detection of internally reflected Cherenkov light (DIRC) detectors~\cite{Kalicy:2020kws},
and electromagnetic and hadronic calorimeters.
The dRICH detector in particular, has already made use of machine learning techniques in its design~\cite{Cisbani:2019xta} and a lot of effort has gone into optimizing the tracking detectors sizes, especially for the lower field BaBar magnet configuration, making use of modern machine learning techniques.   An example of one of these detector configurations is shown in Figure~\ref{fig:ecce}.

Located away from the main detector
are the extremely important auxiliary detectors.
In the far forward direction, a.k.a. the direction of the hadron beam, these detectors detect extremely small angle hadrons and mesons that otherwise would escape the acceptance of the main detector.   
These detectors are key for numerous physics studies including involving the tagging such as the following~\cite{Guzey:2014jva, Jentsch:2021qdp, Friscic:2021oti,Tu:2020ymk}.
In the far backward direction, a.k.a. the direction of the electron beam, small angle electrons can be detected for extremely low $Q^2$ measurements.  Far backward detectors will also be used for  monitoring the luminosity of the EIC experiment similarly to how it was done at HERA~\cite{Helbich:2005qf,ZEUS:2013emt}.

\section{Second Interaction Region}

While the Department of Energy's Electron Ion Collider project only includes plans for one fully instrumented interaction region and detector system, the project nevertheless is preparing for the future by ensuring that a second interaction region and detector can be added.   
As of the writing of this document, it is unclear what the exact timeline of the second region will be, but it is clear that there is a strong focus on making the second region complementary to the first.

To this end, the second interaction region will have a very different far forward region focus and layout.   This is partially driven by the simple fact that the beam crossings at the two different regions must be different simply due to the layout of the EIC accelerator with the electron and ion beam crossing from opposite sides at the two different interaction region locations.   The crossing angle itself will also likely be different with the first interaction region having a crossing angle of 25~mradians while the second interaction region's crossing angle is likely to be 35~mradians.   
For more details on the current ideas for the design of the second interaction region please see~\cite{Gamage:2021wzu}. 

\section{Detector Testing}
\label{testing}

\begin{figure}
    \centering
    \includegraphics[width=\linewidth]{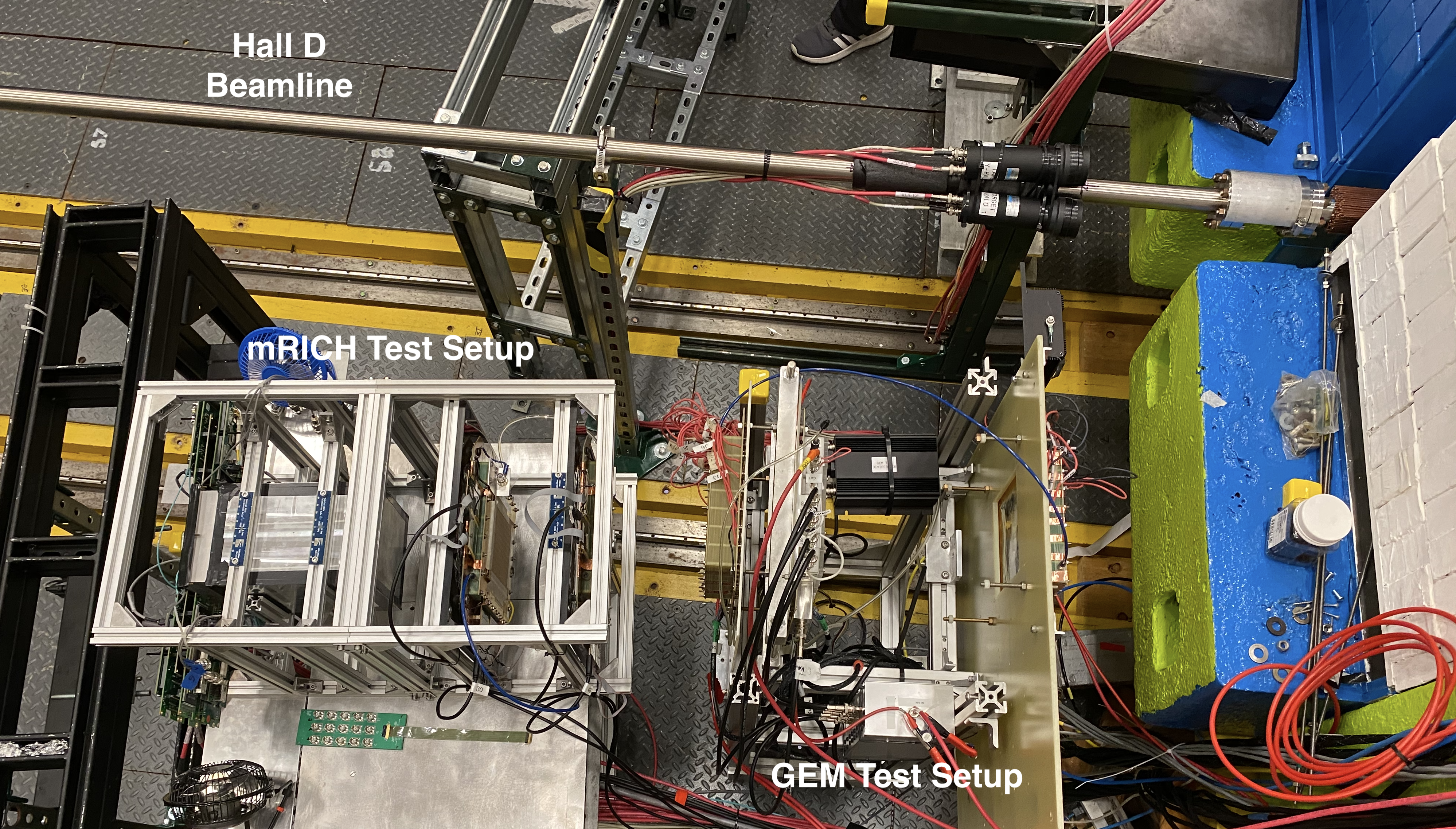}
    \caption{\label{test-setup}Shown is an EIC detector test setup in experimental Hall D at Jefferson Lab.   Electrons from pair production will pass through the GEM and mRICH detectors.    These small scale tests are run in parallel with the main experiment in Hall D, yet provide extremely valuable information about EIC detector performance.}
\end{figure}

An important aspect of preparing a new large scale detector system is the testing of the new detector systems to ensure that they will work as intended.    While many of those tests will be done at traditional detector test locations, detector test are also now underway at Jefferson Lab.   In particular, making use of the electron-positron pair production setup in experimental Hall D.   The area allows for numerous small tests to be conducted parasitically with the main Hall D measurements.   Recent tests include  transition radiation detectors, GEMs, and modular ring imaging Cherenkov detectors.   A photo of one such recent time is shown in Figure~\ref{test-setup}.    The EIC detector testing programs ensure that the detectors will perform as intended and often allow for innovations which further enhance detector performance which otherwise would not have been possible.

\section{Summary}
\label{summary}

It is a very exciting time for the Electron Ion Collider as the project has passed the Department of Energy's critical decision one milestone.  Proto-collaborations are now working to finalize the design of the project detector in preparation for critical decision two.    Presently three groups are working on the possible detector designs: ATHENA, CORE, and ECCE.    While these three designs differ in the details, they are all capable of delivering the science as outlined by the National Academy of Science report as well as well as noted in the EIC white paper and yellow report.
In 2022, the design for the project detector will be finalized, while in parallel work will continue on design for a complementary second detector to be added to the EIC facility in the future.   All of these detectors will benefit greatly from AI and machine learning optimizations.

\acknowledgments

 This work was supported by the U.S. Department of Energy contract DE-AC05-06OR23177 under which Jefferson Science Associates operates the Thomas Jefferson National Accelerator Facility.

\bibliographystyle{JHEP}
\bibliography{mybib}

\end{document}